\documentclass{ifacconf}
\usepackage{multirow}
\usepackage{placeins}
\usepackage{graphicx}      
\usepackage{natbib}        
\usepackage{amsmath}
\usepackage{xurl} 

\usepackage{enumerate}
\usepackage{tikz}
\usetikzlibrary{shapes.geometric, arrows}
\usetikzlibrary{shapes,arrows}
\usepackage{float}
\usepackage{xfrac}
\usepackage{graphics} 
\usepackage{tabularx} 
\usepackage{graphicx}
\usepackage{epstopdf}
\usepackage{dblfloatfix}
\usepackage{epsfig} 
\usepackage{mathptmx} 
\usepackage{times} 
\usepackage{amsmath} 
\usepackage{amssymb}  
\usepackage{comment}
\usepackage{caption}
\usepackage{mathtools}
\DeclarePairedDelimiter\abs{\lvert}{\rvert}
\makeatletter
\let\oldabs\abs
\def\abs{\@ifstar{\oldabs}{\oldabs*}}
\begin{document}
\begin{frontmatter}
\vspace{-15pt}
\title{\small This work has been submitted to IFAC for possible publication}
\vspace{-15pt}
\title{Generalized two-point visual control model of human steering for accurate state estimation} 

\author[First]{Rene E. Mai}
\author[First]{Katherine Sears}
\author[First]{Grace Roessling}
\author[First]{Agung Julius}
\author[First]{Sandipan Mishra}
\address[First]{Rensselaer Polytechnic Institute, 
   Troy, NY 12180 USA (e-mail: mair@rpi.edu, searsk@rpi.edu, roessg@rpi.edu, agung@ecse.rpi.edu, mishrs2@rpi.edu).}

\begin{abstract}                

We derive and validate a generalization of the two-point visual control model, an accepted cognitive science model for human steering behavior. The generalized model is needed as current steering models are either insufficiently accurate or too complex for online state estimation. We demonstrate that the generalized model replicates specific human steering behavior with high precision (85\% reduction in modeling error) and integrate this model into a human-as-advisor framework where human steering inputs are used for state estimation. As a benchmark study, we use this framework to decipher ambiguous lane markings represented by biased lateral position measurements. We demonstrate that, with the generalized model, the state estimator can accurately estimate the true vehicle state, providing lateral state estimates with under 0.25 m error on average across participants. However, without the generalized model, the estimator cannot accurately estimate the vehicle's lateral state. 
\end{abstract}


\end{frontmatter}

\section{Introduction}
\vspace{-3pt}

Although most drivers in the United States learn to drive before they are legally considered adults, driving is a far more complex task than this fact suggests. Humans underestimating the complexity of driving can lead to fatalities \citep{national_center_for_statistics_and_analysis_state_2023}. Experts across the world hope to reduce fatalities through increasing adoption of autonomous and shared autonomous technology like lane-keep assist and adaptive cruise control. Still, even these simple systems can be confused by something as innocuous as ambiguous lane markings \citep{Peiris2022, Shaw2018}. In fact, an incorrect interpretation of the lane marking can lead to dramatic--and dangerous--behavior changes \citep{Boloor2019}. Humans are generally not confused by the sorts of lane markings that stymie autonomous vehicles, so blending human perception with autonomous precision in lane-centering it would be advantageous and increase safety. 

Our goal is therefore to build a state estimation scheme that combines driver steering inputs with more traditional autonomous sensors such as lane-centering cameras and speedometers. For this, we propose using human driver input in the form of human steering commands, in conjunction with sensors to estimate the autonomous vehicle's state; however, this requires an accurate human steering model that fits into existing state estimation frameworks such as Kalman filtering. Further, the human steering model should be resilient to different driving situations such as lane-centering, changing lanes, and negotiating curves, as well as across different speeds and drivers.

Unfortunately, although many researchers have studied human driving behavior, no single model describes all human steering behavior. The two-point visual control model or ``two-point model", is versatile enough to be applicable to many driving scenarios such as lane-changing, curve negotiation, and vehicle following \citep{salvucci_two-point_2004}. The two-point model describes human steering as a proportional-integral control scheme based on visual angles that are functions of the vehicle's lateral position and velocity. Researchers have improved this model in a variety of ways, including adding delays replicating human sensory delays, intermittent attention, and anticipatory control behavior \citep{lappi_visuomotor_2018}. Other efforts have modified the two-point model by developing tuning strategies to update control gains across different latencies and vehicle speeds without human data \citep{mirinejad_modeling_2018}. Human steering can also be modeled as a linear quadratic regulator, again operating on visual angles and the vehicle's lateral position and velocity \citep{nash_simulation_2022}. Recently, more complex data-driven steering models have been proposed that use a variety of sensors such as cameras, microphones, and EEGs \citep{negash_driver_2023}. Although there is rich literature on driver modeling, existing studies repeatedly note that these models require speed- and driver-specific parameter tuning or training \citep{salvucci_two-point_2004, nash_identification_2020-1, ortiz_characterizing_2022, negash_driver_2023}.

We propose and validate a general human steering model based on the two-point model, which we call the ``generalized model." The generalized model replicates human steering behavior with high accuracy \textbf{without} speed or driver-dependent gain tuning, yet is simple enough to implement as part of an extended Kalman filter estimating vehicle state. We also use the generalized model for state estimation via the human-as-advisor architecture introduced in prior work \citep{mai_human-as-advisor_2023}. 

\begin{figure}[h]
    \centering
    \includegraphics[width=0.45\textwidth]{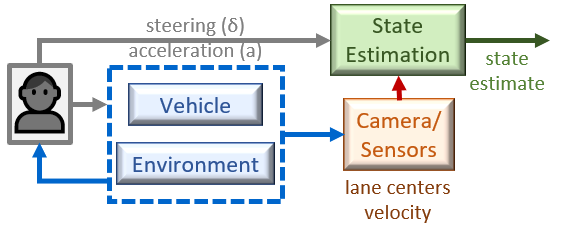}
    \vspace{-4pt}
    \caption{\textit{State estimation with human-as-advisor.}}
    \label{fig: human as advisor}
\vspace{-4pt}
\end{figure}

\begin{figure}[h]
    \centering
    \includegraphics[width=0.45\textwidth]{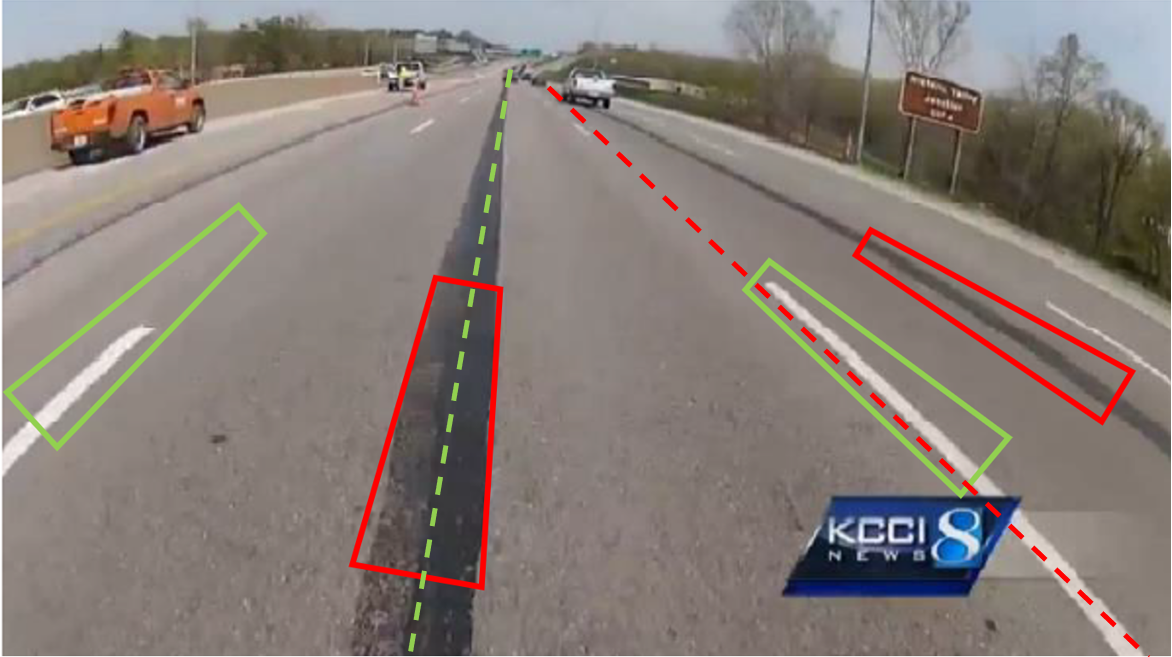}
    \caption{\textit{Ambiguous lane center, adapted from \cite{Shaw2018}.}}
    \label{fig: unclear lane center}
\end{figure}

In this letter, the human controls the vehicle directly; measurements of the vehicle state and human steering input are used for state estimation. The resulting architecture is shown in Fig. \ref{fig: human as advisor}.
This letter makes two key improvements over prior work: (1) developing and validating a generalized model relating human steering inputs to the vehicle state, and (2) designing and demonstrating a state-estimation strategy using the generalized model. We do this by collecting steering trajectories from 9 participants. These trajectories are used for state estimation in a vehicle faced with an ambiguous lane center as in Fig. \ref{fig: unclear lane center}. When provided with with human steering input interpreted through the generalized model, the state estimator accurately determines the vehicle's position (within 0.25 m).

\vspace{-3pt}

\section{Vehicle and sensor model}
\begin{figure*}
    \centering
    \vspace{3mm}\setlength\belowcaptionskip{-1.5\baselineskip}
\includegraphics[width=.99\textwidth]{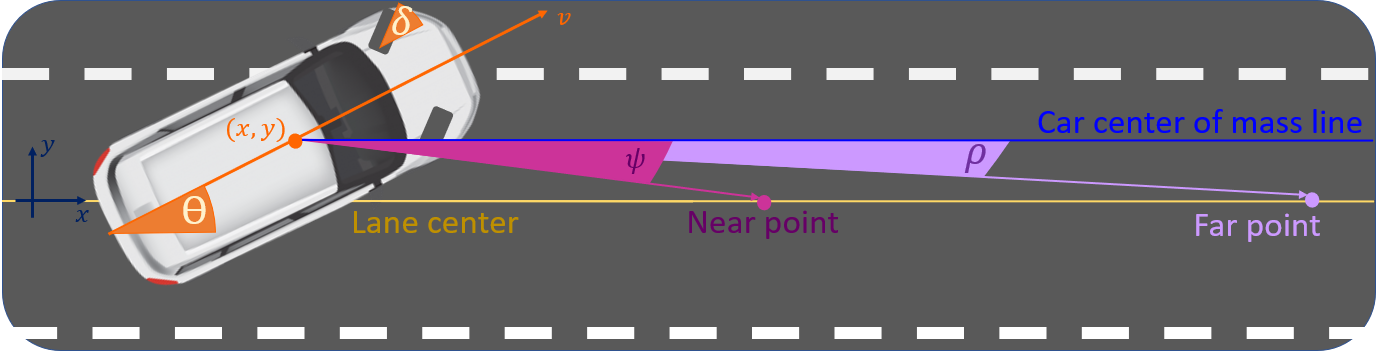}
    \caption{\textit{Near-point and far-point illustration with kinematic bicycle model}}
\vspace{-2mm}
    \label{fig:near-point far-point with kinematic bicycle model}
\end{figure*}

\vspace{-3pt}
Participants and state estimation use a vehicle modeled on the rear-wheel bicycle kinematic model, shown in Fig. \ref{fig:near-point far-point with kinematic bicycle model}.
The vehicle dynamics are given by:
\begin{eqnarray}
\label{eqn_NonlinearContState}
\xi = \begin{bmatrix} \dot{x} \\ \dot{y} \\ \dot{v} \\ \dot{\theta}
\end{bmatrix} = 
\begin{bmatrix} v \cos(\theta) \\ v \sin(\theta) \\ a \\ v \kappa \tan(\delta)
\end{bmatrix} \triangleq \begin{bmatrix} \text{longitudinal velocity} \\ \text{lateral velocity} \\ \text{acceleration} \\ \text{yaw rate} 
\end{bmatrix}.
\end{eqnarray}
The driver commands a steering angle $\delta$ and acceleration $a$. These control inputs are related to the second derivative of the vehicle's longitudinal and lateral position, $\ddot x$ and $\ddot y$, as follows:
\begin{eqnarray}  \label{eqn_FeedbackLin}
\begin{bmatrix}
\ddot x \\
\ddot y%
\end{bmatrix}
\triangleq \underset{\triangleq R(v,\theta )}{\underbrace{%
\begin{bmatrix}
\cos (\theta ) & -v^{2}\kappa \sin (\theta ) \\
\sin (\theta ) & v^{2}\kappa \cos (\theta )%
\end{bmatrix}%
}}%
\begin{bmatrix}
a \\
\tan (\delta )%
\end{bmatrix}.
\end{eqnarray}
 The mapping between the non-linear ($a$, $\tan \delta$) and the linear ($\ddot x$, $\ddot y$) control inputs is invertible as long as $v \neq 0$. This is sensible, as the car is only controllable when moving.

\textit{Lane ambiguity model} We model the uncertain lane center shown in Fig. \ref{fig: unclear lane center} as a bias state that cannot be controlled or directly observed without human input. We discretize Eqs. \eqref{eqn_NonlinearContState}-\eqref{eqn_FeedbackLin} with a sampling time $T_s$ and model the possible lane centers as a Gaussian mixture model, with each mixture component representing a possible lane center. At each time step $k$, the vehicle has a number of possible lane centers $i$; thus, the $i$th component at time $k$ represents the vehicle's state if the lane center corresponding to component $i$ is the true lane center.  Each component has a likelihood of being correct; the likelihoods sum to 1.  Thus, we represent each GMM component as 
\begin{equation}
    \mathcal{G}_i(k) \sim \mathcal{N}(\hat \xi_i(k), \hat \Sigma_i(k), w_i(k)), i \in \{0, 1, 2, ... N\} \text{ where}
\end{equation}
$N$ is the number of components in the Gaussian mixture.
 
 Adding in the bias as a state, the discrete-time vehicle model is
 \begin{eqnarray}
 \label{eq: disc time closed loop}
\xi(k+1) = \underset{\triangleq A}{\underbrace{\left[ \begin{array}{cccc}
         1 & 0 & 0 & 0 \\
         0 & 1 & T_s & 0 \\
         0 & 0 & 1 & 0 \\
         0 & 0 & 0 & 1
     \end{array} \right]}}\xi(k)+\underset{\triangleq B}{\underbrace{\left[ \begin{array}{cc}
         T_s & 0  \\
         0 & \frac{T_s^2}{2} \\
         0 & T_s \\
         0 & 0
     \end{array}\right]}}\begin{bmatrix}
         \ddot x(k) \\ \ddot y(k)
     \end{bmatrix}, 
 \end{eqnarray}  
where $ \left[\begin{array}{cccc}
          \xi_1  &
          \xi_2 &
          \xi_3 &
          \xi_4
     \end{array}\right]^T = \left[\begin{array}{cccc}
          \dot x  &
          y &
          \dot y &
          b
\end{array}\right]^T$. We assume the autonomous vehicle has two sensors: a speedometer measuring longitudinal velocity and a camera measuring the lateral position within the lane. The camera can only detect lateral position to a constant bias, which represents the offset due to erroneous lane markings. Mathematically, the autonomous sensor model is $z_1 = \xi_1+ w_1, \ \ z_2 = \xi_2+ \xi_4+ w_2,$ where $w_{1,2}$ are zero-mean Gaussian measurement noise. Because the vehicle is feedback linearized, $\ddot x$ and $\ddot y$ are not available for control; the vehicle must be controlled the vehicle through $\delta$ and $a$. As in the standard kinematic Kalman filter \citep{jeon_benefits_2007}, we treat the measurement of $\ddot x$ and $\ddot y$ as the inputs to our plant \eqref{eq: disc time closed loop}.
\vspace{-4pt}

\section{Human steering model}
\vspace{-4pt}

As noted in the Introduction, there are several human steering models, many of which begin from the same basic information: the vehicle's lateral position and yaw angle. We begin with the two-point visual control model \citep{salvucci_two-point_2004}. The two-point model is simple enough to use in state estimation via Kalman filtering \textit{and} versatile enough to adapt to a variety of driving tasks.  Further, the two-point model's visual angles are the building blocks for almost every state-of-the-art human steering model, including prominent data-based steering models \citep{nash_simulation_2022, nash_identification_2020-1, Tuhkanen2019, lappi_visuomotor_2018, zhao_human-like_2023}.
\vspace{-6pt}

\subsection{Two-point visual control model} 
\vspace{-3pt}

In this section we discuss the two-point model as it relates to the work of \cite{salvucci_two-point_2004}.
The two-point model uses two visual angles, $\phi$ (the ``near-point angle") and $\Omega$ (the ``far-point angle") to model human steering behavior. $\phi$ and $\Omega$ are the angles between the current direction of the vehicle, $\theta$, and the near-point and far-point distances, respectively, as shown in Fig. \ref{fig:near-point far-point with kinematic bicycle model}. 
The near-point angle is defined as $\theta + \psi$ and the far-point angle as $\theta + \Omega$, where $\psi$ and $\Omega$ are the angles formed by the deviation of the vehicle's center of mass from the lane center and the near- and far-point distances (6.2 and 10-20 m in front of the vehicle, respectively). The far-point angle adjusts steering behavior for long-term road curvature, while the near-point angle accounts for immediate lane centering. 

Therefore, we can express these visual angles as
\begin{eqnarray}
\vspace{-3pt}
    \label{eq: phi basic derivation}
    \phi \triangleq\theta + \psi = \tan^{-1}\frac{\xi_3}{\xi_1} + \tan^{-1}\frac{\xi_1}{6.2},\\
    \Omega \triangleq \theta + \rho = \tan^{-1}\frac{\xi_3}{\xi_1} + \tan^{-1}\frac{\xi_2}{20}.
\end{eqnarray}
The two-point model incorporates $\phi$ and $\Omega$ into a proportional-integral controller. The resulting model can be written in autoregressive discrete-time form as
\begin{multline}
    \delta(k) \!=\! \delta(k\!-\!1)\! +\! \sum_{i=0}^1 k_n\phi(k\!-\!i)\!+\!\sum_{i=0}^1 k_f\Omega(k\!-\!i)\!+\!k_iT_s\phi(k).
\end{multline}
This model produces human-like steering behavior, and is fit to individual's steering behavior by adjusting $k_n$, $k_f$, and $k_i$.
\vspace{-4pt}
\subsection{Generalized two-point visual control model}
\vspace{-5pt}

\label{sec: generalized two point model}
We developed a generalized model because the two-point model requires driver-dependent \citep{salvucci_two-point_2004} and potentially speed-dependent \citep{ortiz_characterizing_2022} tuning. The generalized model is an autoregressive model of the form
\begin{multline}
\label{eq: general steering model}
    \!\!\delta(k)\!\! = \!\!\sum_i \!\!a_i \delta(k\!\!-\!\!i)\! + \!\sum_i\!\! b_i \phi(k\!\!-\!\!i)\!\! +\!\! \sum_i\!\! c_i \Omega(k\!\!-\!\!i) \!\!+ \!\! \sum_i\!\! d_i \xi_3(k\!\!-\!\!i).
\end{multline}
This modification allows the predicted steering angle to better account for the vehicle's second-order dynamics by allowing more autoregressive terms. The $\xi_3$ terms account for vehicle lateral momentum, and the increased memory for $\phi$ makes steering smooth and repetitive--human-like. We identified the correct order for each regressive term (\textit{see} Eq. \eqref{eq: fitted general model}) by searching from the lowest order for each term, increasing the order until there was no significant prediction improvement\footnote{We are also performing studies to very that the chosen model order optimizes the Bayesian information criterion and is no larger than necessary.}. The residual analysis in Sec. \ref{sec: steering model results} shows this model order is sufficient to capture all dynamics represented in the steering input. 
\vspace{-5pt}

\subsection{Steering model validation results}
\vspace{-5pt}

\label{sec: steering model results}
We tested the generalized and two-point models on two tracks: a straight road (9 drivers) and a curved track (1 driver). We fit both models to a single steering trajectory on the straight road from the driver who completed both tracks, the ``illustrative participant" or ``Ill." The resulting two-point model is: 
\begin{equation}
    \label{eq: fitted Salvucci and Gray model}
    \!\!\delta(k)\! =\! \delta(k\!\!-\!\!1)\!\!+\!\!\sum_{i=0}^1\!\!-\!\!0.2 \phi(k\!-\!i)\!\!+\!\!\sum_{i=0}^1 \!\!0.7 \Omega(k\!\!-\!\!i)\!\!+\!\!0.087T_s \phi(k).
\end{equation}
For the generalized model, we first determined the correct order and then fit the model to the same steering trajectory used to fit the model in Eq. \eqref{eq: fitted Salvucci and Gray model}. The resulting generalized model is
\begin{multline}
    \label{eq: fitted general model}
    \!\!\delta(k)\!\! =\!\! \sum_{i=1}^2\!\! a_i \delta(k\!-\!i) \!\!+\!\! \sum_{i=0}^3 \!\!b_i \phi(k\!-\!i) \!\!+\!\! c_0 \Omega(k) \!+\! \sum_{i=0}^1\!\! d_i \xi_3(k\!-\!i),
\end{multline}
where $a_0 = 1.47$, $a_1 = 0.51$, $b_0 = -5.73$, $b_1 = 17.32$, $b_2 = -17.65$, $b_3 = 6.12$, $c_0 = 0.11$, $d_0 = 0.02$, and $d_1 = -0.02$.
\vspace{-3pt}
\subsubsection{Straight road results}
\begin{figure}[h]
    \centering
    \includegraphics[width = 0.45\textwidth]{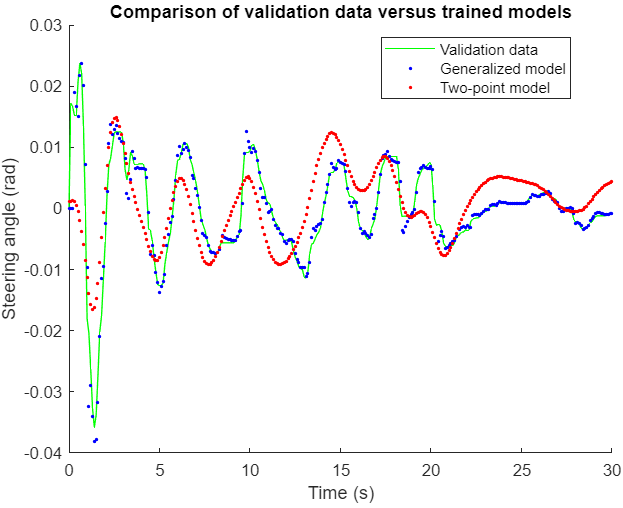}
    \caption{\textit{Steering performance comparison on validation data.}}
    \label{fig: steering model fit}
    \vspace{-4pt}
\end{figure}

Fig. \ref{fig: steering model fit} compares the steering trajectory from a validation run to the predicted steering angles from the two-point and the generalized models. Visually, the generalized model outperforms the two-point model; the error measures agree with this visual assessment. 

 Table \ref{tab:steering accuracy comparison} contains the results from applying the fitted two-point and generalized models to each of the 9 drivers. Although not visible in the data, the drivers had very different behavior during the trials (\textit{see} Sec. \ref{sec: state est results} for more details). Here, again, the generalized model (``Gen") outperforms the two-point model. The generalized model has an average 85\% lower root mean square error (``RMSE Dec."), and the coefficient of determination (``$R^2$ acc") for each driver indicates the generalized model is a good predictor of steering behavior, despite their differing speeds.
\begin{table}[h]
    \centering
    \begin{tabular}{|c|c|c|c|c|c|}
    \hline
    \multirow{2}{2.75em}{\textbf{Driver}} & \multicolumn{2}{c|}{\textbf{Two-point model}} & \multicolumn{2}{c|}{\textbf{Generalized model}} &\multirow{2}{3em}{\textbf{RMSE Dec.}}\\ \cline{2-5}
     & \textit{RMSE (rad)} & \textit{$R^2$ Acc} & \textit{RMSE (rad)} & \textit{$R^2$ Acc} &  \\
    \hline
    Ill. & 0.005 & 0.084 & $7\!\!\times\!\!10^{-4}$ & 0.974 & 87\% \\
    1 & 0.006 & 0.091 & $5\!\!\times\!\!10^{-4}$ & 0.9311 & 92\% \\
    2 & 0.0082 & 0.047 & 0.0014 & 0.942 & 82\% \\
    3 & 0.0076 & 0.007 & $5\!\!\times\!\!10^{-4}$ & 0.926 & 94\% \\
    4 & 0.0033 & -0.033 & $9\!\!\times\!\!10^{-4}$ & 0.865 & 73\% \\
    5 & 0.0055 & 0.062 & $4\!\!\times\!\!10^{-4}$ & 0.904 & 92\% \\
    6 & 0.004 & -0.063 & $9\!\!\times\!\!10^{-4}$ & 0.938 & 74\% \\
    7 & 0.0087 & -0.022 & $7\!\!\times\!\!10^{-4}$ & 0.967 & 92\% \\
    8 & 0.0095 & 0.032 & 0.0023 & 0.939 & 75\% \\
    \hline
    \end{tabular}
    \caption{Average steering angle root mean squared error (``RMSE") and coefficient of determination.}
    \label{tab:steering accuracy comparison}
    \vspace{-4pt}
\end{table}

\begin{figure}[h]
\vspace{-4pt}
    \centering
\includegraphics[width=0.48\textwidth]{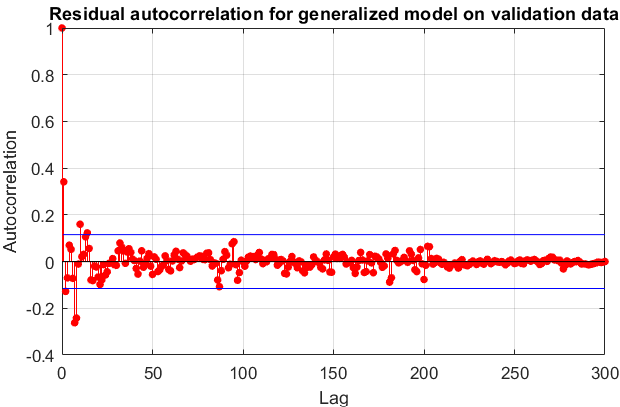}
    \caption{\textit{Autocorrelation of $\delta_h-\hat \delta_h$ for the generalized model.}}
    \label{fig: steering model white noise}
\end{figure}

Further, the generalized model residuals on the training and validation datasets are white, as shown by the autocorrelation of the model error shown in Fig. \ref{fig: steering model white noise}. The generalized model, therefore, captures the full human steering dynamics\footnote{The residuals of the best-fit two-point model on even the training data are not white, reinforcing the fact it does not fully capture human steering dynamics.}. 

\vspace{-3pt}
\subsubsection{Curved road results}
\label{sec: curved road}
\begin{figure}[h]
    \centering
    \includegraphics[width=0.45\textwidth]{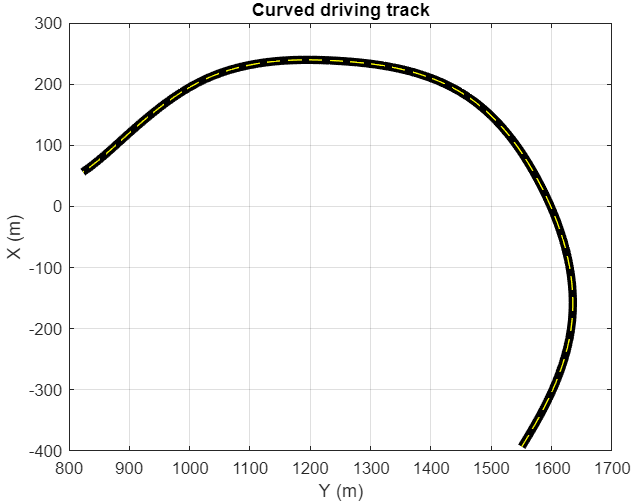}
    \caption{\textit{The curved track the illustrative participant drove along.}}
    \label{fig: curved track}
\end{figure}

\begin{figure}
    \centering
    \includegraphics[width=0.45\textwidth]{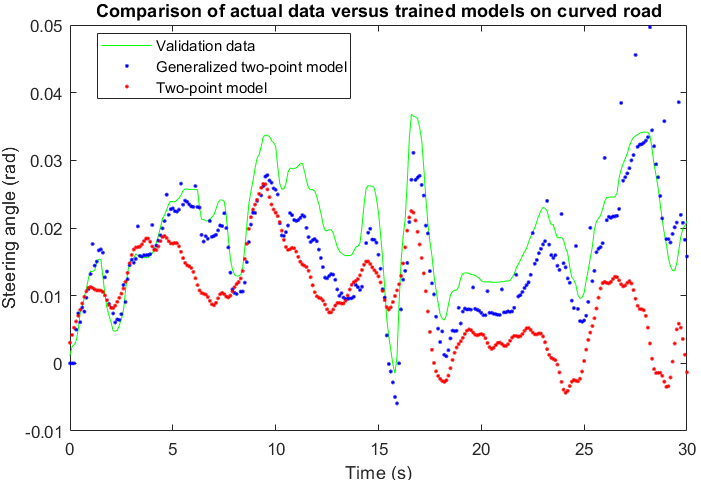}
    \caption{\textit{Steering performance comparison on curved road.}}
    \label{fig: curved road steering}
\end{figure}

The generalized model \textit{also} predicts the human steering input on a curved road well, such as the track shown in Fig. \ref{fig: curved track}. The illustrative participant drove along this track; the resulting steering trajectory is shown in Fig. \ref{fig: curved road steering}, along with the predictions of the driver's steering input based on the two-point and generalized two-point steering model.

As Fig. \ref{fig: curved road steering} clearly shows, the generalized model outperforms the two-point model. The error values for the two models reflect this fact as well. The coefficient of determination for each model's performance on the track was $\begin{bmatrix}
        R^2_{gen} &
        R^2_{two-point}
    \end{bmatrix} = \begin{bmatrix}
        0.665 &-0.141
    \end{bmatrix},$ while the RMSE for each model on this same track was $\begin{bmatrix}
        RMS_{gen} &
        RMS_{two-point} 
    \end{bmatrix} = \begin{bmatrix}
        0.089 & 0.546
    \end{bmatrix},$ which represents an 84\% decrease in RMSE. 

Thus, the generalized model in Sec. \ref{sec: generalized two point model} \textbf{predicts human steering trajectories accurately across participants and speeds, as well as on a strongly curved road}, despite being trained on a \textbf{single} 30 second trajectory from a \textbf{single} participant.
\vspace{-2pt}
\section{Vehicle state estimation using human steering input}
\vspace{-2pt}
\subsection{Overall State-Space Model for Estimation}
\vspace{-3pt}
We incorporate the fitted steering model of Eq. \eqref{eq: fitted general model} by adding each time step backwards as a new state. This yields the following net closed-loop system
\begin{eqnarray}
\label{eqn: discrete full nonlinear prop}
\xi (k+1) =\underset{\triangleq F(\xi(k))}{\underbrace{\begin{bmatrix}
    A_{D} \xi (k) \\
    g_1(\xi(k)) \\
    g_2(\xi(k)) \\
    g_3(\xi(k))
\end{bmatrix}}}+B_{D}\begin{bmatrix}
    \ddot x(k) \\ \ddot y(k)
\end{bmatrix}+\nu(k), \text{ where} 
\end{eqnarray}
$\nu(k)$ is zero-mean white controller noise and 
\noindent\begin{align*}\!\!\!\left[ \!
\begin{array}{c}
     \!g_1(\!\xi\!)  \\
     \!g_2(\!\xi\!) \\
     \!g_3(\!\xi\!)
\end{array}\!\!\right]\!\!\! =\!\!\! \left[\!\!\! \begin{array}{c}
 b_3 \phi \\
(\!b_2\!+\!a_2b_0\!) \phi\!+\!a_2c_0 \Omega\!+\! a_2d_0 \xi_3+ a_2g_1(\!\xi\!)+g_3(\!\xi\!) \\
\!(\!b_1\!+\!a_1b_0\!)\phi \!+\! (\!d_1\!+\!a_1d_0\!) \xi_3\!+\! a_1 c_0 \Omega \!+\! a_1 g_1(\!\xi\!)\!+\!g_2(\!\xi\!)
\end{array}
\!\!\!\right].
\end{align*}
Further, $A_D\triangleq \left[
\begin{array}{cc}
A & 0_{4x3}
\end{array}%
\right] ,~B_{D}\triangleq \left[
\begin{array}{c}
B\\
0_{3x2}
\end{array}%
\right]$.%

Including human input $\delta_h$, the discrete-time sensor model is
\begin{align}
    \label{eq: discrete full nonlinear sensor}
    &\eta(k) = h(\xi(k),\omega(k))\nonumber \\
            &\triangleq \begin{bmatrix}
        \xi_{1}(k)+ \omega_1(k)\\
        \xi_{2}(k)+\xi_{4}(k)+ \omega_2(k)\\
        g_3(\xi(k))+b_0 \phi(k) + c_0 \Omega(k) + d_0 \xi_3(k) + \omega_3(k) \\
    \end{bmatrix},
\end{align}
where $\omega_i(k)$ is zero-mean white noise for each sensor.

We linearize the state transition matrix of Eq. \eqref{eqn: discrete full nonlinear prop}, yielding
\begin{equation}
    \label{eq: linear state transition}
    \frac{\partial F}{\partial \xi} \!=\! \underset{\triangleq \mathcal{A}_h}{\underbrace{
    \left[\begin{array}{ccccccc}
1 & 0 & 0 & 0 & 0 & 0 & 0\\ 
    0& 1& T_s& 0& 0& 0 & 0\\ 
    0& 0& 1& 0& 0& 0 & 0 \\
    0 & 0 & 0 & 1 & 0 & 0 & 0 \\
    0 & \frac{b_3}{6.2} &\frac{b_3}{\xi_1}& 0 & 0 & 0 & 0 \\
    0 & \frac{b_2+a_2b_0}{6.2}\!+\!\frac{a_2c_0}{6.2} &\frac{b_2+a_2b_0+a_2c_0}{\xi_1}\!+\!a_2d_0 &0& 1& 0& a_2\\
    0 & \frac{b_1+a_1b_0}{6.2}\!+\!\frac{a_1c_0}{20}&
    \frac{b_1+a_1b_0+a_1c_0}{\xi_1}\!+\!d_1\!+\!a_1d_0 &0&
    0 & 1& a_1
    \end{array}
    \right]
    }}
\end{equation}
and linearizing the sensor in Eq. \eqref{eq: discrete full nonlinear sensor} likewise yields
\begin{equation}
    \label{eq: linear sensor}
    \frac{\partial h}{\partial \xi} \!=\! \underset{\triangleq \mathcal{C}_h}{\underbrace{
    \left[\begin{array}{ccccccc}
    1 & 0 & 0 & 0 & 0 & 0 & 0 \\ 
    0& 1& 0& 1& 0& 0 & 0 \\ 
   0& \frac{b_0}{6.2}+\frac{c_0}{20}& \frac{b_0+c_0}{\xi_1}+d_0& 0&0& 0& 1
    \end{array}
    \right].
    }}
\end{equation}

We can easily see that $\left(\mathcal{A}_h, B_D\right)$ is not controllable, as the lane-centering bias $\xi_4$ cannot be controlled. Further, suppose we define the first two rows of $\mathcal{C}_h$ as $C$. In that case, we can see that $\left(\mathcal{A}_h, C\right)$ is unobservable as the lateral position and bias states are only observed together; however, $\left(\mathcal{A}_h, \mathcal{C}_h\right)$ is observable so long as \textit{any} of the $b_i$ or $c_0$ are non-zero. 
\vspace{-3pt}
\subsection{State estimation via EKF with a Gaussian mixture model}
\label{sec: state estimation}
\vspace{-4pt}

In this section we discuss the usage of state estimation as in \cite{mai_human-as-advisor_2023}. We use an extended Kalman filter to estimate the states $\xi_i$, obtaining state estimate $\hat \xi_i$, with a GMM as described in  representing the multiple possible lane centers. The Gaussian mixture extended Kalman filter propagation occurs as detailed in the literature. Thus, when a new measurement $\eta$ arrives, the estimated state $\hat \xi_i$ is updated as
\begin{eqnarray}
\label{eq: GMM EKF correction}
\rho_{i,k} = \mathcal{C}_h \hat \Sigma_{i, k \vert k-1} \mathcal{C}_h^T + R, \\
L_{i,k} = \hat \Sigma_{i,k \vert k-1} \mathcal{C}_h^T \rho_{i,k}^{-1}, \\
\hat \xi_{i,k \vert k} = \xi_{i,k\vert k-1} + L_{i,k}(\eta_k - \hat \eta_{i,k\vert k-1}), \\
\hat \Sigma_{i,k \vert k} = (I-L_{i,k}\mathcal{C}_h)\hat \Sigma_{i,k \vert k-1}.
\end{eqnarray}
where $\hat \eta_{i,k\vert k-1}$ is the expected measurement as described in Eq. \eqref{eq: discrete full nonlinear sensor} assuming the $i$th component is correct.

The components $\hat \xi_{i}$ are weighted based on each component's probability of being true based on the residual between the actual measurement and the expected measurement, $\Tilde{z}_k = \eta_k - \hat \eta_{i,k\vert k-1}$. Each component's weight is therefore
\begin{eqnarray}
\label{eqn_Weighting}
    \bar w_{i, k} = \frac{w_{i, k-1}}{2\pi \sqrt{\rho_{i,k}}}exp\frac{-\frac{1}{2} \Tilde{z}_{i,k}^T\Tilde{z}_{i,k}}{\rho_{i,k}}
    \textrm{ where } w_{i\vert k} = \frac{\bar w_{i\vert k}}{\sum_{i=1}^{N} \bar w_{i\vert k}}
\end{eqnarray} and the aggregate weighted state estimate $\hat \xi_k = \sum_{i=1}^{N}\hat \xi_{i, k \vert k}w_{i, k}$ is determined at each time step.
\vspace{-3pt}
\subsection{State estimation results}
\label{sec: state est results}
\vspace{-3pt}
We collected data from participants using the Simulink Highway Lane Following Toolbox. The toolbox and its capabilities are described in more detail in \cite{mai_human-as-advisor_2023}. Participants controlled the vehicle directly through a Logitech G29 steering wheel and pedal set. They were instructed the vehicle would start off-center in the right-hand lane and were asked to center the vehicle in that lane and told to control vehicle speed as they wished, so long as they did not stop the vehicle; as a result, some participants reached speeds above 45 m/s (more than 100 mph), while others maintained a constant speed. If requested, users were given one trial run or a demonstration. Each participant completed 10 runs lasting 30 seconds each; the illustrative participant completed 14 runs on the straight track. We collected the applied steering and acceleration/braking commands, as well as the vehicle states in the simulator.

We used the extended Kalman filter described in Sec. \ref{sec: state estimation} to estimate the vehicle's true state (represented by the states collected from the simulator). The vehicle started at $\xi_0 = \left[\begin{array}{ccc}
     \dot x& y & \dot y 
\end{array}\right] = \left[\begin{array}{ccc}
     15 \text{ m/s}& -0.5 \text{ m} & 0 \text{ m/s} 
\end{array}\right]$ in each simulation. The extended Kalman filter is formulated so the state estimator is faced with two equally likely lane centers offset by 1.8 m (the half-lane width). The state begins with an identical initial error to that faced by the human drivers. Thus, the initial Gaussian mixture model for the vehicle states is
\begin{eqnarray}
\vspace{-3pt}
\label{eq: gaussianmixturedef}
\!\!\mathcal{G}_1 \sim \mathcal{N} (\xi_1, I_4, 0.5), 
\mathcal{G}_2 \sim \mathcal{N} (\xi_2, I_4, 0.5), \  \text{where} 
\end{eqnarray} 
\begin{equation*}
    \xi_1 = \!\begin{bmatrix}
    15 & -0.5 & 0 & 0
    \end{bmatrix}^T, \xi_2 = \!\begin{bmatrix}
   15 & 1.3 & 0 & -1.8
    \end{bmatrix}^T.
\end{equation*}
The initial values of the starting states, $\xi_{i,5}$, $\xi_{i,6}$, and $\xi_{i,7}$ are driver-dependent, but are identical for both components as they are human-observed values and not affected by camera bias.
\vspace{-3pt}
\subsubsection{Illustrative results for one participant}
\label{sec: illustrative results}
\begin{figure}[h]
    \centering
    \includegraphics[width =0.45\textwidth]{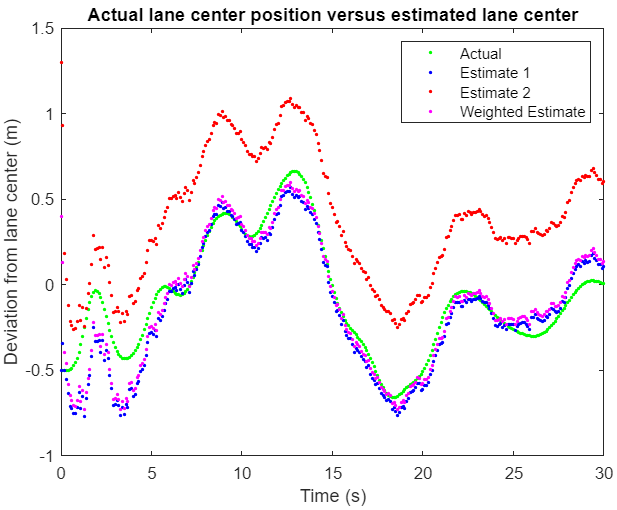}
    \caption{\textit{Lateral position estimates for the validation run in Fig. \ref{fig: steering model fit}.}}
    \label{fig: validation run lateral estimate}
\end{figure}

Here, we analyze results from the simulation runs performed by the illustrative participant. Fig. \ref{fig: validation run lateral estimate} shows the lateral state estimation trajectory for one of the validation runs performed by this participant.

\begin{figure}[b]
    \centering
    \includegraphics[width = 0.45\textwidth]{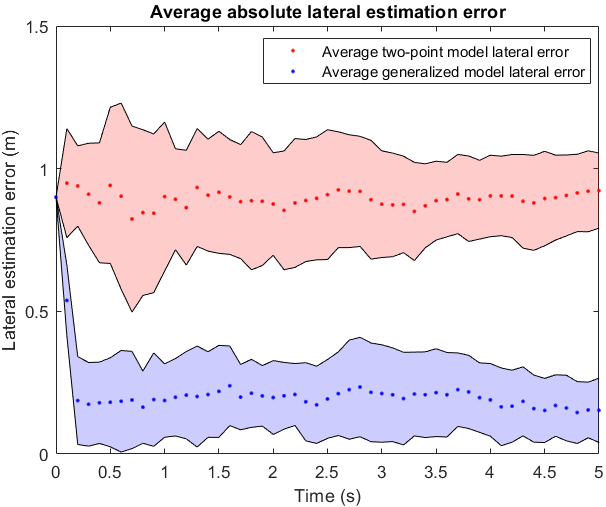}
    \caption{\textit{Lateral error evolution for the illustrative participant for the generalized and the two-point model.}}
    \label{fig: lateral error analysis}
\end{figure}

Fig. \ref{fig: lateral error analysis} shows the average lateral estimation error for the illustrative participant over the first 5 seconds of each run, with the standard deviation of the error superimposed. Using the generalized model, the lateral estimation error takes only 3 time steps to reach its steady state value of about 0.2 m; without the generalized model, the error is much higher, indicating inability to discern between the true and illusory lanes. 

Thus, the generalized model, combined with the state estimator described in Sec. \ref{sec: state estimation}, provides \textbf{accurate state estimation even in the face of an unobservable bias}.
\vspace{-3pt}
\subsubsection{Tabulated results across all participants}

We include results for 8 additional participants along with the illustrative results in Sec. \ref{sec: illustrative results} above. Each participant completed the same course, with their data used in the same extended Kalman filter as the illustrative participant. Below, we tabulate the average root-mean-square lateral position error for each participant across all runs. We also provide comparative results to an extended Kalman filter based on the two-point model. As might be expected, the estimation error results are much better for the generalized model, since it predicts steering trajectories much more accurately. Note that, without human input, the mean error is (as might be expected) 0.9 m, which reflects perfect confusion between the two lane centers.

\begin{table}[h]

    \centering
    \begin{tabular}{|c|c|c|c|c|}\hline
    \multirow{2}{2.75em}{\textbf{Driver}} & \multicolumn{2}{c|}{\textbf{Two-point model}} & \multicolumn{2}{c|}{\textbf{Generalized model}} \\ \cline{2-5}
     & \textit{Mean (m)} & \textit{STD ($m^2$)} & \textit{Mean (m)} & \textit{STD ($m^2$)} \\
    \hline
        Ill & 1.0 & 0.06 & 0.22 & 0.11 \\
        1 & 1.0 & 0.11 & 0.18 & 0.04 \\
        2 & 0.95 & 0.11 & 0.21 & 0.10 \\
        3 & 0.97 & 0.18 & 0.25 & 0.15 \\
        4 & 0.98 & 0.17 & 0.18 & 0.06 \\
        5 & 0.95 & 0.11 & 0.17 & 0.06 \\
        6 & 1.0  & 0.15 & 0.18 & 0.05 \\
        7 & 0.93 & 0.17 & 0.25 & 0.15 \\
        8 & 0.96 & 0.11 & 0.18 & 0.06 \\ \hline
    \end{tabular}
    \caption{Average lateral RMSE (mean and standard deviation) for all participants and all runs.}
    \label{tab: lateral RMSE for all participants}
\end{table}

Interestingly, although both the generalized model and the two-point model were trained on the illustrative participant (\textit{see} Sec. \ref{sec: illustrative results}), the results for the other participants show similar performance for \textit{both} models. This implies that driver-dependent tuning may be less important than initially believed. However, the generalized model significantly outperforms the two-point model, with a maximum lateral RMSE across \textit{all} participants of 0.25 m (under one foot). 

Thus, we have shown that the generalized model, combined with the state estimator in Sec. \ref{sec: state estimation}, provides accurate state estimation in the face of an unobservable bias \textbf{across multiple participants with different driving styles and drive speeds.}
\vspace{-3pt}
\section{Conclusion}
\vspace{-3pt}

This paper explored state estimation via human-as-advisor, using two steering models: an accepted cognitive science model, the two-point model, and the generalized two-point model. We demonstrated that the generalized model accurately predicts human input, providing an average 85\% reduction in modeling error across participants and speeds, on both a straight and a curving track. We further demonstrated that state estimation with the generalized model is accurate, while the estimator using the two-point model accurately estimate the vehicle's state. Future work includes verifying that: (1) the generalized model's fitted order optimizes the Bayesian information criterion; (2) the generalized model provides a good estimate when the vehicle is autonomously driven, rather than under full human control; and (3) expanding tests to include multiple participants and a variety of track shapes.

\bibliography{ifacconf} 
\bibliographystyle{ifacconf}
                                                   







\end{document}